# Adatbiztonsági protokollok teljesítmény analízise

# Performance Evaluation of Security Protocols

# Evaluarea performanţelor protocoalelor de securitate

*dr. GENGE Béla[1], dr. HALLER Piroska[2]*

[1,2] "Petru Maior" Egyetem, Marosvásárhely, ROMÁNIA
{[1]bgenge, [2]phaller}@upm.ro

*Abstract*
*We propose a comparative performance evaluation of security protocols. The novelty of our approach lies in the use of a polynomial mathematical model that captures the performance of classes of cryptographic algorithms instead of capturing the performance of each algorithm separately, approach that is used in other papers. A major advantage of using such a model is that it does not require implementation-specific information, because the decision is based on comparing the estimated performances of protocols instead of actually evaluating them. The approach is validated by comparatively evaluating the performances of 1000 automatically generated security protocols against the performances of their actual implementations.*

*Abstract*
*În cadrul acestei lucrări propunem o metodă comparativă de evaluare a performanţelor protocoalelor de securitate. Noutatea metodei constă în utilizarea unui model matematic polinomial ce capturează performanţa claselor de algoritmi criptografici, faţă de capturarea performanţei fiecărui algoritm în parte, după cum s-a procedat în alte lucrări. Un avantaj major al acestui model este că nu necesită informaţii legate de implementarea protocoalelor, întrucât decizia este luată prin compararea performanţelor estimate şi nu prin compararea performanţelor măsurate. Metoda propusă este validată prin evaluarea comparativă a 1000 de protocoale de securitate generate automat, raportat la performanţele măsurate ale acestora.*

*Összefoglaló*
*A dolgozatban egy új teljesítmény analízis módszert javasolunk adatbiztonsági protokollokra. Az eddig javasolt módszerek a megvalósítások teljesítményét mérték és nem vizsgálták külön az egyes algoritmusokat. A módszer újdonsága abban áll, hogy a protokollokban használt algoritmusok teljesítményét egy polinomiális közelítő függvény segítségével írjuk le a bennük szereplő paraméterek függvényében. A javasolt eljárás fő előnye, hogy nem szükséges információkat tudni a megvalósításról az algoritmusok vagy paramétereik kiválasztásánál, mert a döntés a becsült, összehasonlított teljesítmények alapján történik. A javasolt módszert 1000 generált adatbiztonsági protokoll teljesítmény analízisével érvényesítettük, összehasonlítva a megvalósított protokollok lemért teljesítményével.*

## 1. BEVEZETŐ

Adatbiztonsági protokolloknak nevezzük azokat a kommunikációs protokollokat, melyek célja, hogy az adatokat kizárólag a protokoll részvevői érhessék el. Ezek a protokollok a kriptográfiára alapoznak, mely segítségével az üzeneteket titkosítani és hitelesíteni lehet. Ezért az adatbiztonsági protokollok teljesítmény analízisét a kriptográfiai algoritmusok teljesítmény analízisével kell kezdeni. Az említett algoritmusok teljesítményét számos tényező befolyásolyhatja, mint a használt kódolási módszer, az üzenet mérete, a kódolási kulcsok mérete, valamint a fizikai rendszer paraméterei (pl. processzor, memória).

A kriptográfiai algoritmusok teljesítményének a méréséhez használni lehet az energia fogyasztást [1, 2] vagy az algoritmusok végrehajtási idejét [3], de a jelen dolgozatban a végrehajtási időt használtuk a teljesítmények összehasonlítására.

A szakirodalomban a legtöbb teljesítmény analízis módszer a kriptografiai algoritmusok esetén a megvalósítások elemzésére alapoz [4, 5, 6, 7, 8], míg a modell-alapú módszerek sokkal ritkábbak [9, 10]. A modell-alapú módszerek is tartalmaznak megvalósítás függő paramétereket, ezért a teljesítmény analízis eredményeit csak bizonyos rendszereken lehet használni.

Az általunk javasolt, modell-alapú teljesítmény analízis módszer nem tartalmaz megvalósítás-specifikus adatokat. Ezért használni lehet rendszer tervezéskor, az adatbiztonsági protokoll kiválasztására, mikor a megvalósítástól függő paraméterek még nem állnak rendelkezésre.

A módszer az adatbiztonsági protokollok informális leírásából indul ki és kiszámítja azok becsült teljesítményét figyelembe véve a használt kriptográfiai algoritmusok teljesítmény modelljét. Mivel a rendszer paraméterei teljesen hiányoznak, a kidolgozott teljesítmény modell számításba kell vegye algoritmusok összes megvalósítási lehetőségét. A kidolgozott teljesítmény modellt összehasonlítjuk a megvalósított protokollok mért teljesítményével, bizonyítva azok helyességét.

## 2. KRIPTOGRÁFIAI ALGORITMUSOK TELJESITMÉNY MODELLJE

A kriptográfiai algoritmusok teljesítményének elemzéséhez a három legismertebb kriptográfiai könyvtárat használtuk: *Cryptlib* [11], *OpenSSL* [12] és *Crypto++* [13]. Mértük az egyes algoritmusok futási idejét változtatva azok paramétereit. Mivel a méréseket többszálas környezetben végeztük a *QueryPerformanceCounter* függvényt használtuk, mely lehetővé teszi az 1 ms-os pontosságú méréseket. A méréseket Windows XP operációs rendszeren végeztük Intel Dual Core 1.8GHz-es processzort és 1GB RAM-ot használva, az eredmények több mérést átlagát jelentik.

### 2.1 Szimmetrikus algoritmusok teljesítmény analízise

A szimmetrikus algoritmusok teljesítményét a következő paraméterek befolyásolják: üzenet méret, kódolásra használt kulcsok, és kódolási módszer: ECB, CBC, CFB, OFB, CTR.

Mind a három kriptográfiai könyvtár nagy számú algoritmust valósít meg, ezek közé tartozik DES, 3DES, AES, IDEA, CAST vagy Blowfish. Minden algoritmus esetén lemértük a végrehajtási időt az összes megengedett paraméter kombinációra, változtatva ugzanakkor a kódolt adat méretét 16-tól 16384 byte-ig. A végrehajtott kombinációk száma 44 *OpenSSL* használatakor, 40 a *Cryptlib* és a *Crypto++* használatakor. A sok kombináció miatt csak egy részét tüntettük fel az 1. ábrán. Mivel a dekódoló algoritmusok teljesítménye hasonló a kódoló algoritmusok teljesítményéhez, az 1. ábrán csak a kódoló algoritmusok teljesítményét mutatjuk be.

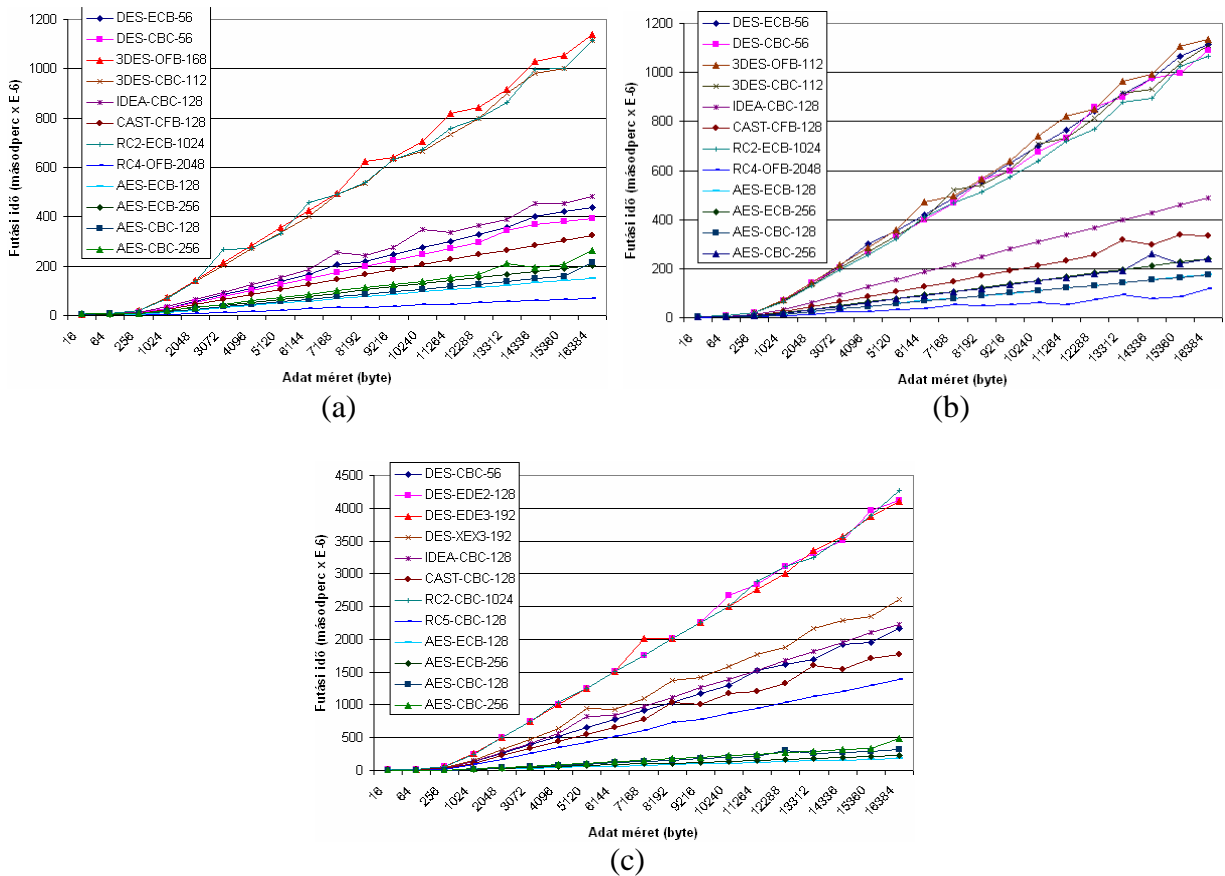

**1. ábra** Szimmetrikus algoritmusok kódolásának futási ideje:
(a) Cryptlib (b) OpenSSL (c) Crypto++

### 2.2 Aszimmetrikus algoritmusok teljesítmény analízise

Az aszimmetrikus algoritmusok két kulcsot használnak föl, egyet a kódolásra és egy másikat a dekódolásra. E algoritmusokat *nyilvános* kulcsú algoritmusoknak is nevezik, mivel, az egyik kulcs nyilvános (publikus), a másik pedig titkos (privát).

A kódolást és dekódolást úgy a nyilvános mint titkos kulcsal lehet elvégezni. Ezek szerint, két típusú aszimmetrikus kódolást ismerünk: az egyik a nyilvános kulcsú kódolás, a másik pedig a titkos kulcsú kódolás.

Az első típusú kódolás a nyilvános kulcsot használja kódolásra és a titkos kulcsot dekódolásra. A második típusú kódolás a titkos kulcsot használja kódolásra és a nyilvános kulcsot dekódolásra. Ezek miatt, az első típust adat kódolásra, míg a másodikat digitális aláírásokra alkalmazzák.

Ahogyan ez a 2. ábrán is látható, a dekódolási művelet sokkal időigényesebb mint a kódolási művelet. A feltüntetett ábrán csak az RSA algoritmust ábrázoltuk, mivel ez az egyedüli megvalósított algoritmus mind a három könyvtárban, mely megenged adatkódolást és digitális aláírást is.

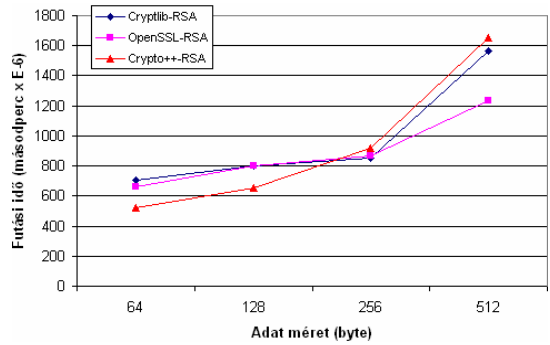
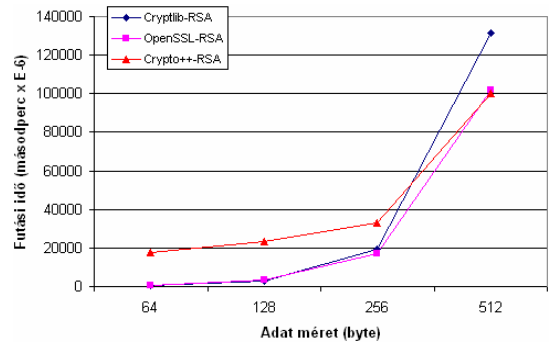

(a)                  (b)

**2. ábra** Aszimmetrikus algoritmusok futási ideje:
(a) Kódolás (b) Dekódolás

### 2.3 Hash algoritmusok teljesítmény analízise

A hash algoritmusok egy tetszőleges hosszú szövegből egy üzenet összefoglalót generálnak, mely fontos szerepet játszik többek között a digitális aláírásokban. Elemzésünkben tanulmányoztuk a három említett könyvtár összes támogatott hash algoritmusát. Ezek közül kiemeljük a következőket: MD4, MD5, SHA-1 és SHA-256, melyek teljesítménye a 3. ábrán látható.

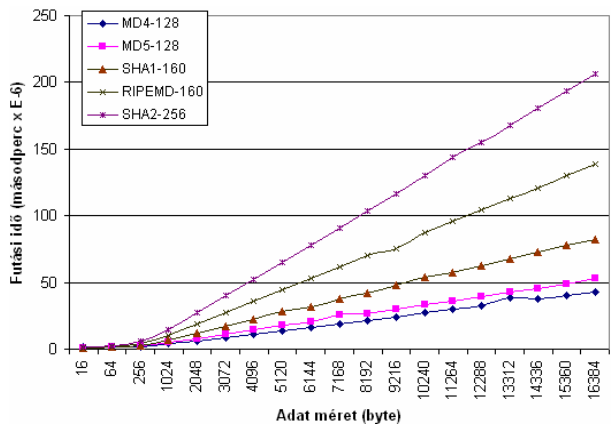
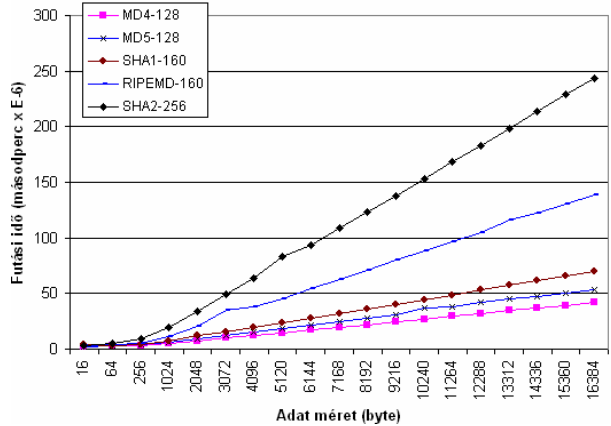

(a)                  (b)

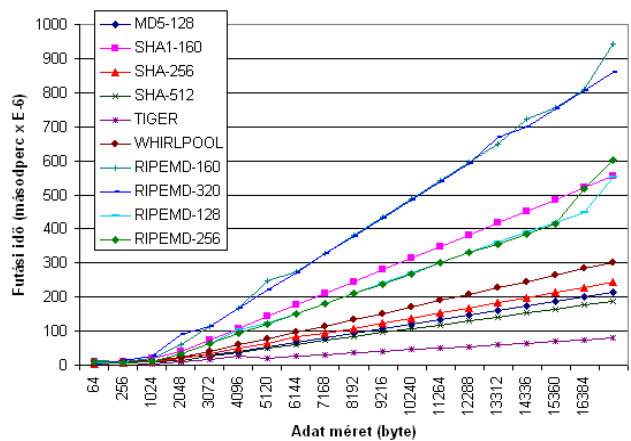

(c)

**3. ábra** Hash algoritmusok futási ideje:
(a) Cryptlib (b) OpenSSL (c) Crypto++

**2.4 A modell felépítése**

Kiindulva az előbb bemutatott mérési eredményekből kidolgoztunk egy polinomiális modellt minden algoritmusra. Alkalmazva a legkisebb négyzetes hiba módszerét [14], minden algoritmus típusra kiszámítottuk a közelítő polinomiális függvény együtthatóinak értékét.

A matematikai modellt a következő harmadfokú polinomiális függvénnyel írjuk le:

$$f(x) = \alpha_4 x^3 + \alpha_3 x^2 + \alpha_2 x + \alpha_1 \tag{1}$$

Az együtthatókat a következő egyenletek segítségével számítottuk ki:

$$\begin{cases} \dfrac{\partial R}{\partial \alpha_1} = -2\sum_{i=1}^{n}\left(y_i - \left(\alpha_4 x_i^3 + \alpha_3 x_i^2 + \alpha_2 x_i + \alpha_1\right)\right) = 0 \\ \dfrac{\partial R}{\partial \alpha_2} = -2x_i\sum_{i=1}^{n}\left(y_i - \left(\alpha_4 x_i^3 + \alpha_3 x_i^2 + \alpha_2 x_i + \alpha_1\right)\right) = 0 \\ \dfrac{\partial R}{\partial \alpha_3} = -2x_i^2\sum_{i=1}^{n}\left(y_i - \left(\alpha_4 x_i^3 + \alpha_3 x_i^2 + \alpha_2 x_i + \alpha_1\right)\right) = 0 \\ \dfrac{\partial R}{\partial \alpha_4} = -2x_i^3\sum_{i=1}^{n}\left(y_i - \left(\alpha_4 x_i^3 + \alpha_3 x_i^2 + \alpha_2 x_i + \alpha_1\right)\right) = 0 \end{cases} \tag{2}$$

A kiszámított együtthatók láthatóak a 1. táblázatban. Használva a mért értékeket és a kidolgozott matematikai modellt, kiszámítottuk a becslési hiba értékét. A kiszámított hiba értéke 3.714 ms, ami a maximális kimért érték 0.3963%-a.

**1. táblázat** Kiszámított együtthatók értéke

| Algoritmus típusa | Művelet | Együtthatók | | | |
|---|---|---|---|---|---|
| | | $\alpha_4$ | $\alpha_3$ | $\alpha_2$ | $\alpha_1$ |
| Szimmetrikus | Kódolás | $6.87723137 \cdot 10^{-13}$ | $-1.158358181 \cdot 10^{-8}$ | 0.05692690466 | 2.6048870112 |
| | Dekódolás | $1.398869423 \cdot 10^{-11}$ | $-2.30355815 \cdot 10^{-7}$ | 0.05710958418 | 2.203380464 |
| Hash | - | $1.522749902 \cdot 10^{-11}$ | $-2.754241881 \cdot 10^{-7}$ | 0.01700037541 | 3.852945249 |
| Aszimmetrikus | Kódolás | $3.594921725 \cdot 10^{-8}$ | -0.00020136227 | 0.47812305376 | 434.828218 |
| | Dekódolás | $1.64038427 \cdot 10^{-6}$ | -0.0019686486 | 6.9193250868 | 3135.53968253 |

# 3. ADATBIZTONSÁGI PROTKOLLOK ANALÍZISE

Felhasználva a kidolgozott modellt összehasonlíthatjuk az adatbiztonsági protokollok teljesítményét ha ezek leírhatók mint az előbbi algoritmusok kombinációi. Első lépésben automatikusan generáltunk 1000 adatbiztonsági protokollt tetszőlegesen kombinálva az algoritmusokat és a paramétereket. Kiindulva az 1000 protokoll-ból, generáltunk 998000 protokoll párat, melyek teljesítményét összehasonlítottuk. Minden protokoll párra kiszámítottuk a becsült és a mért teljesítmény arányt. Az első értéket a modell segítségével számítottuk ki, a második érték pedig a protokollok mért teljesítményének az arányát jelképezi.

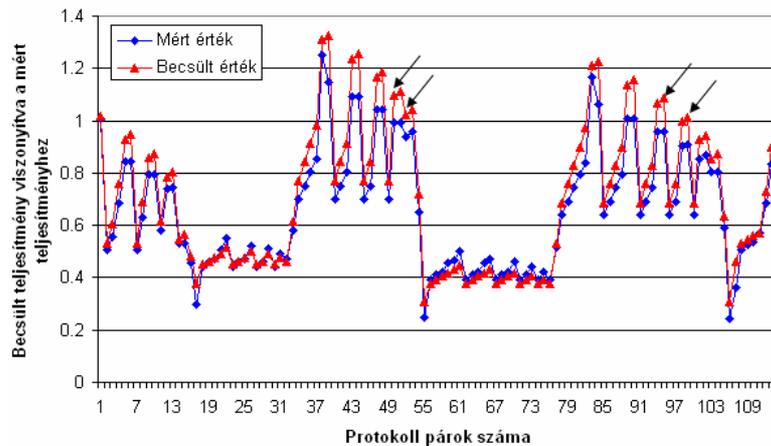

**4. ábra** Lemért és becsült protokoll párok összehasonlítása
AES-CBC, SHA1 és RSA algoritmusokra

Több mérést végeztünk, változtatva több paramétert is a megvalósított protokoll párokra és használva a protokoll modellt a becsült protokollokra. Ahogyan ez a 4. ábrán is látható, a protokollok becsült teljesítménye követi a protokollok mért teljesítményének az értékét.

A nyilakkal feltüntettük azokat az eseteket ahol becslési hiba következett be. A hiba oka legtöbbször az, hogy nagyon rövid protokollok esetén, ahol a futási idő rövidebb mint 1ms a mérési pontosság már nem elegendő. Hasonló a helyzet akkor is ha a két összehasonlított protokoll futási ideje nem tér el legalább 1ms-al.

Mivel a kódolt adat mérete nagymértékben befolyásolja a futási időt tanulmányoztuk ennek hatását a mérési pontosságra. A 4. ábrán a kódolt adatok mérete 80 Byte volt, és ez egy 2.9%-os eltérést generált a bescsült és mért értékek között. Ha növeljük vagy csökkentjük az adatok méretét, változik a hiba százaléka is, amint ez látható a 5. ábrán. A maximális hibát 10 Byte-os adatokra kaptuk (7.12%), míg a minimális hibát 300 Byte-os adatokra (1.85%).

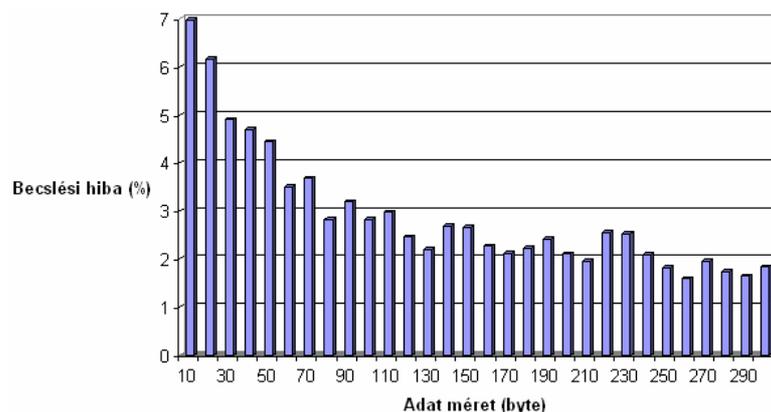

**5. ábra** A becslési hiba változása az adatok méretének függvényében

# 4. KÖVETKEZTETÉSEK

A kidolgozott teljesítmény analízis módszer egy polinomiális modellre épül, melyet három kriptográfiai könyvtár segítségével határoztunk meg: *Cryptlib*, *OpenSSL* és *Crypto++*. Elvégezve az említett könyvtárak által megvalósított kriptográfiai algoritmusok kimerítő teljesítmény analízisét, meghatároztunk egy közelítő polinomiális függvényt az algoritmusok teljesítményére. A modell figyelembe veszi az egyes algoritmusokban szereplő paramétereket. A kidolgozott modellt adatbíztonsági protokollok teljesítményének az elemzésére használtuk fel. Mivel a módszer összehasonlító elvekre alapszik használható protokollok teljesítmény alapú kiválsztására még a tervezési fázisban mikor a megvalósítási paraméterek még nem állnak rendelkezésünkre. A nagy számú kísérlet igazolta, hogy a kidolgozott módszer alkalmazható adatbíztonsági protokollok teljesítményének a becslésére.